\documentclass[12pt]{article}
\usepackage{amsmath}
\usepackage{times}
\usepackage{graphicx}
\usepackage{color}
\usepackage{multirow}
\usepackage[authoryear]{natbib}
\usepackage{rotating}
\usepackage{bbm}
\usepackage{latexsym}

\newcommand{\captionfonts}{\normalsize}

\makeatletter  
\long\def\@makecaption#1#2{%
  \vskip\abovecaptionskip
  \sbox\@tempboxa{{\captionfonts #1: #2}}%
  \ifdim \wd\@tempboxa >\hsize
    {\captionfonts #1: #2\par}
  \else
    \hbox to\hsize{\hfil\box\@tempboxa\hfil}%
  \fi
  \vskip\belowcaptionskip}
\makeatother   

\usepackage{amssymb}
\usepackage{diagrams}

\begin{document}
\hspace{13.9cm}1

\ \vspace{20mm}\\

{\LARGE Parametric inference in the large data limit}

{\LARGE using maximally informative models}

\ \\
{\bf \large Justin B. Kinney $^{\displaystyle 1, \displaystyle 2}$ and Gurinder S. Atwal$^{\displaystyle 1}$}\\
{$^{\displaystyle 1}$Simons Center for Quantitative Biology, Cold Spring Harbor Laboratory,\\ Cold Spring Harbor, NY 11724 }\\
{$^{\displaystyle 2}$ Please direct correspondence to jkinney@cshl.edu.}\\


\thispagestyle{empty}
\markboth{}{NC instructions}
\ \vspace{-0mm}\\
%
\begin{center} {\bf Abstract} \end{center}
Motivated by data-rich experiments in transcriptional regulation and sensory neuroscience, we consider the following general problem in statistical inference. When exposed to a high-dimensional signal $S$, a system of interest computes a representation $R$ of that signal which is then observed through a noisy measurement $M$. From a large number of signals and measurements, we wish to infer the ``filter'' that maps $S$ to $R$. However, the standard method for solving such problems, likelihood-based inference, requires perfect \emph{a priori} knowledge of the ``noise function'' mapping $R$ to $M$. In practice such noise functions are usually known only approximately, if at all, and using an incorrect noise function will typically bias the inferred filter. Here we show that, in the large data limit, this need for a pre-characterized noise function can be circumvented by searching for filters that instead maximize the mutual information $I[M;R]$ between observed measurements and predicted representations. Moreover, if the correct filter lies within the space of filters being explored, maximizing mutual information becomes equivalent to simultaneously maximizing every dependence measure that satisfies the Data Processing Inequality. It is important to note that maximizing mutual information will typically leave a small number of directions in parameter space unconstrained. We term these directions ``diffeomorphic modes'' and present an equation that allows these modes to be derived systematically. The presence of diffeomorphic modes reflects a fundamental and nontrivial substructure within parameter space, one that is obscured by standard likelihood-based inference. 
\section{Introduction}

This paper discusses a familiar problem in statistical inference, but focuses on an under-studied limit that is becoming increasingly relevant in the era of large data sets. Consider an experiment having the following form:
\begin{diagram}[LaTeXeqno]
\underset{\rm signal}{S} & \rTo^{\stackrel{\rm filter}{\theta(S)}} & \underset{\rm representation}{R} & \rTo^{\stackrel{\rm noise~function}{\pi(M|R)}} & \underset{\rm measurement}{M}. \label{eq:srm}
\end{diagram}
When presented with a signal $S$, a system of interest applies a deterministic filter $\theta$ thereby producing an internal representation $R$ of that signal. For each representation $R$, a noisy measurement $M$ is then generated. The conditional probability distribution $\pi(M | R)$ from which $M$ is drawn is called the ``noise function'' of the system. From data consisting of $N$ signal-measurement pairs, $\{S_n, M_n\}_{n=1}^N$, we wish to reconstruct the filter $\theta$. This paper focuses on how to infer $\theta$ properly in the $N \to \infty$ limit when the noise function $\pi$ is unknown \emph{a priori}.

All statistical regression problems have this ``SRM'' form \citep{Bishop2006}, but we will focus on two biological applications for which this problem is particularly relevant. In neuroscience, SRM experiments are commonly used to characterize the response of neurons to stimuli \citep{SchwartzEtAl2006}. For instance, $S$ may be an image to which a retina is exposed, while $M$ is a binary variable (`spike' or `no spike') indicating the response of a single retinal ganglion cell. It is often assumed that the spiking probability depends on a linear projection $R$ of $S$. The specific probability of a spike given $R$ is determined by the noise function $\pi$. 

More recently, analogous experiments have been used to characterize the biophysical mechanisms of transcriptional regulation. In the context of work by \citet{KinneyEtAl2010}, $S$ is the DNA sequence of a transcriptional regulatory region, $R$ is the rate of mRNA transcription produced by this sequence, and $M$ is a (noisy) measurement of the resulting level of gene expression. The filter $\theta$ is a function of DNA sequence that reflects the underlying molecular mechanisms of transcript initiation. The noise function $\pi$ accounts for both biological noise\footnote{Such as stochastic gene expression \citep{ElowitzEtAl2002}.} and instrument noise.  

The standard approach for solving inference problems like these is to adopt a specific noise function $\pi$, then search a space $\Theta$ of possible filters for the one filter $\theta$ that maximizes the likelihood $p(\{M_n\} | \{S_n\}, \theta, \pi) = e^{N L(\theta,\pi)}$, where,
\begin{eqnarray}
L(\theta,\pi) = \frac{1}{N} \sum_{n=1}^N \log \pi(M_n | \theta(S_n)), \label{eq:likelihood}
\end{eqnarray}
is the per-datum log likelihood. For instance, the method of least squares regression corresponds to maximum likelihood inference assuming a homogenous Gaussian noise function $\pi$ \citep{Bishop2006}. 

Although the correct filter $\theta$ does indeed maximize $L(\theta,\pi)$ when the correct noise function $\pi$ is used, full \emph{a priori} knowledge of this noise function is rare in practice. Often $\pi$ is chosen primarily for computational convenience, as is standard with least-squares regression. This can be problematic because using an incorrect $\pi$ will \emph{typically} produce bias in the inferred filter $\theta$, bias that does not disappear in the $N \to \infty$ limit. The reason for this is illustrated in Fig.\ 1.

Sometimes this problem can be partially alleviated by performing a separate ``calibration experiment'' in which the noise function $\pi(M|R)$ is measured directly. For instance, one might be able to make repeated measurements $M$ for a select number of known representations $R$. However, there will always be residual measurement error in $\pi$ that will propagate to $\theta$ in a manner that is not properly accounted for by simply plugging $\pi$ into likelihood calculations via Eq.\ \ref{eq:likelihood}. 

An alternative inference procedure \citep{SharpeeEtAl2004, Paninski2003, KinneyEtAl2007} that circumvents the need for an assumed noise function is to maximize the mutual information \citep{CoverThomas1991}, 
\begin{eqnarray}
I(\theta) = I[R;M] = \int dR\, dM\, p(R,M) \log \frac{p(R,M)}{p(R)p(M)}, \label{eq:mi}
\end{eqnarray}
between predictions $R$ and measurements $M$.\footnote{The notation $I(\theta)$ and $I[R;M]$ will be used interchangeably.} Here, $p(R,M)$ is the empirical joint distribution between predictions and measurements, and thus depends implicitly on $\theta$. This method has been proposed, studied, and applied in the specific contexts of receptive field inference \citep{SharpeeEtAl2004, Paninski2003, SharpeeEtAl2006, PillowSimoncelli2006} and transcriptional regulation \citep{KinneyEtAl2007, ElementoEtAl2007, Kinney2008,  KinneyEtAl2010, MelnikovEtAl2012}. However, this alternative approach can be applied to a much wider range of statistical regression problems, and a general discussion of how maximizing mutual information relates to maximizing likelihood for arbitrary SRM systems has yet to be presented.

We begin by pointing out that, in the $N \to \infty$ limit, maximizing mutual information over $\theta$ alone is equivalent to maximizing likelihood over \emph{both} $\theta$ and $\pi$. We then prove that when the correct filter $\theta$ lies within the class of filters being considered, maximizing mutual information is also equivalent to simultaneously maximizing every dependence measure that satisfies the Data Processing Inequality (DPI). However, in the absence of a known noise function $\pi$, SRM experiments are fundamentally incapable of constraining certain directions in the parameter space of $\theta$; we call these directions ``diffeomorphic modes.'' An equation for diffeomorphic modes is described and then applied to filters having various functional forms. In particular, our analysis of a linear-nonlinear filter used by \citet{KinneyEtAl2010} to model transcriptional regulation demonstrates how model nonlinearities can eliminate diffeomorphic modes in useful and non-obvious ways.  This has important consequences for biophysical studies of transcriptional regulation that use recently developed DNA-sequencing-based assays \citep{KinneyEtAl2010, MelnikovEtAl2012}. 

Throughout this manuscript, $R$ is used to implicitly denote the representation predicted by the filter $\theta$ for signal $S$, i.e.\ $R \equiv \theta(S)$. $\mathcal{D}(\theta) \equiv \mathcal{D}[R;M]$ is used to denote any DPI-satisfying dependence measure.  Representations $R$ are assumed to be multidimensional with components $R^{\mu}$, and $\partial_{\mu} \equiv \partial / \partial R^{\mu}$. $\theta$ is used to denote both a filter and the parameters governing that filter. $\Theta$  represents both an abstract space of filters, as well as the space of parameters for filters assumed to have a specific functional form. In the latter case, $\theta^i$ denotes coordinates in parameter space, and $\partial_i \equiv \partial / \partial \theta^i$. 

\section{Mutual information and likelihood}

We begin by discussing the connection between likelihood and mutual information in the $N \to \infty$ limit. In this limit, the per-datum log likelihood (Eq.\ \ref{eq:likelihood}) can be rewritten as,
\begin{eqnarray}
L(\theta,\pi) &=& \int dR\, dM\, p(R,M) \log \pi(M|R) \label{eq:L} \\
&=& I(\theta) - D(\theta,\pi) - H[M].  \label{eq:decomposition}
\end{eqnarray}
The first term, $I(\theta)$, is the mutual information between $R$ and $M$ (Eq.\ \ref{eq:mi}) and is independent of the noise function $\pi$. The second term,
\begin{eqnarray}
D(\theta, \pi) = \int dR\, dM\, p(R,M) \log \frac{p(M|R)}{\pi(M|R)},
\end{eqnarray}
is the Kullback-Leibler (KL) divergence between the empirical distribution $p(M|R)$, which results from the choice of $\theta$, and the assumed noise function $\pi(M|R)$. The last term, $H[M] = - \int dM\, p(M) \log p(M)$, is the entropy of the measurements $M$. $H[M]$ is independent of both $\theta$ and $\pi$ and can thus be ignored in the optimization problem. 

The key point is that finding maximally informative filters $\theta$ is equivalent to solving the maximum likelihood  problem over \emph{both} filters $\theta$ and noise functions $\pi$. This is because if $\theta$ maximizes $I(\theta)$, simply choosing a noise function that matches the empirical noise function, i.e.\ setting $\pi(M|R) = p(M|R)$, will minimize $D(\theta,\pi)$ and thus maximize $L(\theta, \pi)$. 

If one can formalize prior assumptions about the noise function $\pi$ using a Bayesian prior $p(\pi)$,  the relevant objective function becomes the per-datum marginal likelihood,
\begin{eqnarray}
L_m(\theta) &\equiv& \frac{1}{N} \log \int d\pi\, p(\pi) p(\{M_n\} | \{S_n\}, \theta, \pi). \label{eq:marginal_likelihood}
\end{eqnarray}
This is analogous to Eq.\ \ref{eq:L} computed after all possible noise functions have been integrated out. As has been shown in previous work \citep{KinneyEtAl2007, RajanEtAl2013}, maximizing marginal likelihood and maximizing mutual information are essentially equivalent in the $N \to \infty$ limit. This can be seen by decomposing $L_m(\theta)$ as,
\begin{eqnarray}
L_m(\theta) &=& I(\theta) - \Delta(\theta) - H[M], \label{eq:marginal_log_likelihood}
\end{eqnarray}
where,
\begin{eqnarray}
\Delta(\theta) \equiv -\frac{1}{N} \log \left[ \int d\pi\, p(\pi) e^{- N D(\theta, \pi)} \right].
\end{eqnarray}
Under weak assumptions about the prior $p(\pi)$,\footnote{E.g.\ $p(\pi)$ does not vanish at the true noise function $\pi^*$.} $\Delta \to 0$ as $N \to \infty$ (see Appendix A).


\section{DPI-optimal filters}

Mutual information is just one measure among many that satisfy DPI (see Appendix B). In this section, we discuss the importance of DPI for the SRM inference problem and introduce the notion of ``DPI-optimal'' filters.

\citet{Paninski2003} has argued as follows for using DPI-satisfying dependence measures as objective functions for inferring filters. If $\theta^*$ is the correct filter in an SRM experiment, then for every filter $\theta$, 
\begin{diagram}[LaTeXeqno]
R & \lTo^{\theta}  & S & \rTo^{\theta^*} & R^* & \rTo^{\pi^*} & M, \label{eq:markov}
\end{diagram}
is a Markov chain. This implies $\mathcal{D}(\theta) \leq \mathcal{D}(\theta^*)$ for every DPI-satisfying measure $\mathcal{D}$. If $\theta^*$ resides within the space $\Theta$ of filters being explored, it must therefore fall within the subset of $\Theta_{\mathcal{D}} \subseteq \Theta$ on which $\mathcal{D}$ is maximized. As a simple extension of this argument, we point out that, because $\theta^*$ maximizes \emph{all} DPI-satisfying measures, $\theta^*$ must actually lie within the intersection of all such sets, i.e.,\
\begin{eqnarray}
\theta^* \in \Theta_{\rm DPI} \equiv \bigcap_{\mathcal{D} {\rm~satisfying~DPI}} \Theta_{\mathcal{D}} 
\end{eqnarray}
Filters in $\Theta_{\rm DPI}$ can properly be said to be ``DPI-optimal.'' 

This raises an important question: would optimizing a variety of different measures $\mathcal{D}$, not just mutual information, narrow the search for $\theta^*$? Here we show the answer is `no'; when $\theta^* \in \Theta$, maximizing mutual information is equivalent to simultaneously maximizing every DPI-satisfying measure, i.e., 
\begin{eqnarray}
\Theta_I = \Theta_{\rm DPI}. \label{eq:I_is_DPI}
\end{eqnarray}

To prove this, we first define on the space of all possible filters a weak and strong partial ordering, as well as an equivalence relation. These mathematical structures are a natural consequence of DPI. For any two filters $\theta_1$ and $\theta_2$,\footnote{The subscripts 1 and 2 label two different filters, not two parameters of a single filter.} we write,
\begin{eqnarray}
{\rm weak~ordering:~~~~}\theta_1 \leq \theta_2 ~~~&{\iff}&~~~\mathcal{D}(\theta_1) \leq \mathcal{D}(\theta_2)~~{\rm for~all~} \mathcal{D}, \\
{\rm strong~ordering:~~~~}\theta_1 < \theta_2~~~&{\iff}&~~~ \theta_1 \leq \theta_2 {\rm~~but~not~~} \theta_2 \leq \theta_1, \label{eq:strong} \\ 
{\rm equivalence:~~~~}\theta_1 \simeq \theta_2~~~&{\iff}&~~~ \theta_1 \leq \theta_2 {\rm~~and~~} \theta_2 \leq \theta_1. \label{eq:equivalence}
\end{eqnarray}
Note that $\theta_1 \leq \theta_2$ if $R_1 \leftrightarrow R_2 \leftrightarrow M$ is a Markov chain. 
The set $\Theta_{\rm DPI}$ of DPI-optimal filters is the supremum of $\Theta$ under this partial ordering. The equivalence $\Theta_I = \Theta_{\rm DPI}$, which occurs when $\theta^* \in \Theta$, follows directly from the fact, proven in Appendix C, that $\theta < \theta^*$ implies $I(\theta) < I(\theta^*).$ We note that this is not true for all DPI-satisfying measures. For instance, the trivial measure $\mathcal{D} = 0$ satisfies DPI but reveals no information about whether a given $\theta$ resides in $\Theta_{\rm DPI}$. These results are illustrated in Fig.\ 2.

\section{Diffeomorphic modes}

Whether or not two filters $\theta_1$ and $\theta_2$ satisfy the above equivalence relation (Eq.\ \ref{eq:equivalence}) can depend on the true filter $\theta^*$ and on the specific noise function $\pi^*$ of the SRM experiment. However, certain pairs of filters will satisfy $\theta_1 \simeq \theta_2$ under \emph{all} SRM experiments. We will refer to such pairs of filters as being ``information equivalent.'' In Appendix D we prove that two filters are information equivalent if and only if their predicted representations are related by an invertible transformation. 

As an objective function, mutual information is inherently incapable of distinguishing between information equivalent filters. In practice this means that selecting maximally informative filters from a parametrized set of filters can leave some directions in parameter space unconstrained. Here we term these directions ``diffeomorphic modes.''

The diffeomorphic modes of linear filters have an important and well-recognized consequence in neuroscience: the technique of maximally informative dimensions can identify only the relevant subspace of signal space, not a specific basis within that subspace \citep{SharpeeEtAl2004, Paninski2003, PillowSimoncelli2006}. However, an interesting twist occurs in applications to transcriptional regulation. Here, linear filters are often used to model the sequence-dependent binding energies of proteins to DNA \citep{Stormo2013}. Any mechanistic hypothesis about how DNA-bound proteins interact with one another predicts that the transcription rate will depend on these binding energies in a \emph{specific} nonlinear manner \citep{BintuEtAl2005, Stormo2013}. Such up-front knowledge about the nonlinearities of linear-nonlinear filters can eliminate diffeomorphic modes of the underlying linear filters in useful and non-obvious ways \citep{Kinney2008, KinneyEtAl2010}. 

\subsection{An equation for diffeomorphic modes}

Consider a filter $\theta$, representing a point in $\Theta$, whose parameters $\theta^i$ are infinitesimally transported along a vector field having components $g^i(\theta)$. This yields a new filter $\theta'$ with components $\theta'^i = \theta^i + \epsilon g^i(\theta)$. If the representation $R$ predicted by $\theta$ for a specified signal $S$ has components $R^{\mu}$ in representation space, these will be transformed to $R'^{\mu} = R^{\mu} + \epsilon \sum_i g^i(\theta) \partial_i R^{\mu}$. 

If the vector field $g^i(\theta)$ represents a diffeomorphic mode of $\Theta$, this transformation must be invertible, meaning the values $\sum_i g^i(\theta) \partial_i R^{\mu}$ cannot depend on $S$ except through the value of $R$. This is a nontrivial condition because $\partial_i R$ can depend on the underlying signal $S$ in an arbitrary manner. However, if  $\sum_i g^i(\theta) \partial_i R^{\mu}$ does indeed depend only on the value of $R$ then,
\begin{eqnarray}
\sum_i g^i(\theta) \partial_i R^{\mu}  = h^{\mu}(R, \theta).\label{eq:diff_mode_def}
\end{eqnarray}
for some vector function $h^{\mu}(R, \theta)$. This is the equation that any diffeomorphic mode $g^i(\theta)$ must satisfy.

\subsection{General linear filters}

We now use Eq.\ \ref{eq:diff_mode_def} to derive the diffeomorphic modes of general linear filters. By definition, a linear filter $\theta$ yields a representation $R$ that is a linear combination of signal ``features'' $F_i^{\mu}$, i.e.,\
\begin{eqnarray}
R^{\mu} = \sum_i \theta^i F_i^{\mu}(S). \label{eq:linear_model}
\end{eqnarray}
As is standard with regression problems \citep{Bishop2006}, the term ``linear'' describes how $R$ depends on the parameters $\theta^i$; the features $F_i^{\mu}$ need not be linear functions of $S$. 

To find the diffeomorphic modes of these filters, we apply the operator $\sum_i g^i(\theta) \partial_i$ to both sides of Eq.\ \ref{eq:linear_model}. Using Eq.\ \ref{eq:diff_mode_def} we then find $
\sum_i g^i(\theta) F_i^{\mu}(S) = h^{\mu}( \{ \sum_j \theta^j F_j^{\nu}(S) \}, \theta).$ The left-hand side is linear in signal features, so unless something unusual happens,\footnote{E.g.\ if the various features $F_i^{\mu}(S)$ exhibit complicated interdependencies, either because of their functional form or because signals $S$ are restricted to a particular subspace. We ignore such possibilities here.} $h^{\mu}(R,\theta)$ must also be a linear function of $R$, i.e.\ have the form,
\begin{eqnarray}
h^{\mu}(R,\theta) = a^{\mu}(\theta) + \sum_\nu b_{\nu}^{\mu}(\theta) R^{\nu}. \label{eq:linear_mode}
\end{eqnarray}

The number of diffeomorphic modes is bounded above by the number of independent parameters on which $h^{\mu}$ depends (at each $\theta$).\footnote{Technically the number of diffeomorphic modes is the number of independent vector fields $g^i$ that correspond to such transformations. However, here we consider only \emph{proper} diffeomorphic modes, not gauge transformations; as in physics, we define gauge transformations to be vector fields $g^i$ along which transformation of $\theta$ leaves all predicted representations invariant.}  For a general linear filter we see that there can be no more than $\dim(R)[\dim(R)+1]$ diffeomoprhic modes, which is the number of parameters $a^{\mu}$ and $b^{\mu}_{\nu}$ in Eq.\ \ref{eq:linear_mode}. This bound is independent of the number of signal features, i.e.\ the dimensionality of $S$. In particular, if $R$ is a scalar, then $h = a + b R$. In this case we observe two diffeomorphic modes, corresponding to additive and multiplicative transformations of $R$. 

\subsection{A linear-nonlinear filter}

\citet{KinneyEtAl2010} performed experiments probing the biophysical mechanism of transcriptional regulation at the \emph{Escherichia coli} \emph{lac} promoter (Fig.\ 3A). These experiments are of the SRM form where $S$ is the DNA sequence of a mutated \emph{lac} promoter, $M$ is a measurement of the resulting gene expression, and the mRNA transcription rate $T$ is the internal representation the system. Linear filters were used to model the binding energies $Q$ and $P$ of the two proteins CRP and RNAP. The specific parametric form used for these filters was,
\begin{eqnarray}
Q = \sum_{bl} \theta_{Q}^{bl} S_{bl} + \theta^0_Q,~~~~~P = \sum_{bl} \theta_{P}^{bl} S_{bl} + \theta^0_P,\label{eq:linear_filters}
\end{eqnarray}
where $b$ indexes the four possible bases (A,C,G,T), $l$ indexes nucleotide positions within the 75 bp promoter DNA region, $S_{bl} = 1$ if base $b$ occurs at position $l$ and $S_{bl} = 0$ otherwise.\footnote{To fix the gauge freedoms of these filters, \citet{KinneyEtAl2010} adopted the convention that $\min_b \theta^{bl}_Q = \min_b \theta^{bl}_P = 0$ for all positions $l$.}

Measurements $M$ were taken for $\sim 5\times10^4$ mutant \emph{lac} promoters $S$. These data were then used to fit a model for the sequence-dependent binding energy of CRP. This was done by maximizing $I[Q;M]$. Because of the diffeomorphic modes of $Q$, the parameters $\theta_{Q}^{bl}$ were inferred up to an unknown scale and the additive constant $\theta^0_Q$ was left undetermined. This is shown in Fig.\ 3B. Analogous results were obtained for RNAP (Fig.\ 3C). 

Next, a full thermodynamic model of transcriptional regulation was proposed and fit to the data. Based on the hypothesized biophysical mechanism, the transcription rate $T$ was assumed to depend on $S$ via,
\begin{eqnarray}
T = \frac{1}{1 + R^{-1}}~~~{\rm where}~~~R = e^{-P} \frac{1 + e^{-Q - \gamma}}{1 + e^{-Q}}. \label{eq:kinney_model}
\end{eqnarray}
This quantity $R$ is called the ``regulation factor'' of the promoter \citep{BintuEtAl2005}. Because $R$ is an invertible function of $T$, it serves equally well as the representation of the SRM system. In the following analysis we work with $R$ instead of $T$ due to its simpler functional form. 

When the parameters of the linear filters $P$ and $Q$ were \emph{simultaneously} fit to data by maximizing $I[T;M]$ (or equivalently, maximizing $I[R;M]$), three of the four diffeomorphic modes described above were eliminated (Fig.\ 3D). Specifically, the overall scale of the parameters $\theta^{bl}_Q$ and $\theta^{bl}_P$ were fixed, allowing binding energy predictions for CRP and RNAP in physically meaningful units of $k_B T$. The parameter $\theta^0_Q$, corresponding to the intracellular concentration of CRP, was also fixed by the data. The only diffeomorphic mode left unbroken was $\theta^0_R$, corresponding to the intracellular concentration of RNAP. 

We now show how the nonlinearity in $R$ was able to break three of the four diffeomorphic modes of $P$ and $Q$.  First observe that any diffeomorphic mode of a linear-nonlinear filter must also be a diffeomorphic mode of \emph{each individual} linear filter if, as here, the linear filters are independent functions of $S$. This means any diffeomorphic mode $g^i$ of the full thermodynamic model for $R$ must satisfy,
\begin{eqnarray}
\sum_i g^i \partial_i R = h = (a_P + b_P P) \partial_{P} R + (a_Q + b_Q Q) \partial_{Q} R + a_{\gamma} \partial_{\gamma} R,
\end{eqnarray}
for coefficients $a_P, b_P, a_Q, b_Q, a_{\gamma}$ which do not depend on $S$. Evaluating the right-hand side derivatives and substituting for $P$ in terms of $Q$ and $R$ we find, 
\begin{eqnarray}
h = - R \left[ a_P - b_P \log \left\{  \frac{R(1 + e^{-Q})}{1+e^{-Q-\gamma}} \right\} -  \frac{(a_Q + b_Q Q)e^{-Q}(1 - e^{-\gamma})}{(1 + e^{-Q- \gamma})(1+e^{-Q})} + \frac{a_{\gamma}  e^{-Q-\gamma}}{1 + e^{-Q-\gamma}} \right].
\end{eqnarray}
For $g^i$ to be a diffeomorphic mode, the right-hand side must be independent of $S$ for fixed $R$. The terms dependent on $Q$ must therefore vanish, rendering $b_P = a_Q = b_Q = a_{\gamma} = 0$.\footnote{This assumes $\gamma \neq 0$, i.e.\ that CRP actually interacts with RNAP. Which is true.} Any diffeomorphic modes $g^i$ must therefore satisfy $\sum_i g^i \partial_i R = -a_P R$. Thus only one mode remains, corresponding to an additive shift in the binding energy $P$. 


\section{Discussion}

Likelihood-based inference masks the fundamentally different ways in which data constrain the parameters that lie along diffeomorphic modes versus those that lie along nondiffeomorphic modes. Standard likelihood inference constrains all model parameters, including both diffeomorphic and nondiffeomorphic modes, with error bars that scale as $N^{-1/2}$.\footnote{In this discussion we ignore gauge parameters, which do not alter model predictions and are therefore non-identifiable.} These constraints will be consistent with the correct underlying filter $\theta^*$ when the correct noise function is used (Fig.\ 4A). However, use of an incorrect noise function will typically cause $\theta^*$ to fall outside the  error bars inferred along \emph{both} diffeomorphic and nondiffeomorphic modes (Fig.\ 4B). 

This problem is rectified if we use a prior $p(\pi)$ that reflects our uncertainty about what the true noise function is. From Eq.\ \ref{eq:marginal_log_likelihood} it can be seen that using the resulting marginal likelihood to compute a posterior distribution on $\theta$ will constrain diffeomorphic and nondiffeomorphic modes in fundamentally different ways (Fig.\ 4C). Nondiffeomorphic modes will be constrained by $I(\theta)$, which remains finite in the large $N$ limit. This produces error bars on nondiffeomorphic modes comparable to those produced by likelihood when the correct noise function $\pi^*$ is used. However, constraints along diffeomorphic modes will come only from $\Delta$. Because $\Delta$ vanishes as $N^{-1}$,\footnote{More precisely, given any direction $i$ in filter space, $\partial_i^2 \Delta|_{\theta^*} \sim N^{-1}$ for $N$ large enough.} diffeomorphic constraints become independent of $N$ once $N$ is sufficiently large. 

Fortunately, one does not need to posit a specific prior probability over all possible noise functions in order to confidently infer filters from SRM data. Using mutual information as an objective function instead of likelihood, i.e.\ sampling filters according to $p(\theta | {\rm data}) \sim e^{NI(\theta)}$, will constrain nondiffeomorphic modes the same way that marginal likelihood does while putting no constraints along diffeomorphic modes (Fig.\ 4D). 

One might worry that a large fraction of filter parameters will be diffeomorphic, and that the analysis of SRM experiments will require an assumed noise function in order to obtain useful results even if doing so yields unreliable error bars. Such situations are conceivable, but in practice this is often not the case. We have shown that for linear filters, the number of diffeomorphic modes will typically not exceed $\dim(R) [\dim(R) + 1]$ regardless of how large $\dim(S)$ is. Some of these diffeomorphic modes may also be eliminated if these linear filters are combined using a  nonlinearity of known functional form. Indeed, of the 204 independent parameters comprising the biophysical model of transcriptional regulation inferred by \citet{KinneyEtAl2010}, only one was diffeomorphic. 

A bigger concern, perhaps, is the practical difficulty of using mutual information as an objective function. Specifically, it remains unclear how to compute $I(\theta)$ rapidly and reliably enough to confidently sample from $p(\theta | {\rm data}) \sim e^{N I(\theta)}$.  Still, various methods for estimating mutual information are available \citep{Khan:2007p1045, Panzeri:2007p911}, and the information optimization problem has been successfully implemented using a variety of techniques \citep{SharpeeEtAl2004, SharpeeEtAl2006, KinneyEtAl2007, KinneyEtAl2010, MelnikovEtAl2012}. We believe the exciting applications of mutual-information-based inference provide compelling motivation for making progress on these practical issues. 

\section{Appendix A: marginal likelihood}

In certain cases $\Delta(\theta)$ can be computed explicitly and thereby be shown to vanish \citep{KinneyEtAl2007}. More generally, when $\pi$ is taken to be finite-dimensional, a saddle-point computation (valid for large $N$) gives $\Delta(\theta) \approx  \frac{1}{2N} {\rm Tr}[ \log \partial \partial \tilde{D}] + {\rm const}.$ Here, $\partial \partial \tilde{D}$ is the $\pi$-space Hessian of $\tilde{D}(\theta,\pi) \equiv D(\theta,\pi) - \frac{1}{N} \log p(\pi)$ computed using $\pi(M|R) = p(M|R)$. If $\log p(\pi)$ and its derivatives are bounded, then the $\theta$-dependent part of $\Delta(\theta)$ decays as $N^{-1}$. If $\pi$ is infinite dimensional, this saddle-point computation becomes a semiclassical computation in field theory akin to the density estimation problem studied by \citet{BialekEtAl1996}. If this field theory is properly formulated through an appropriate choice of $p(\pi)$, then $\Delta(\theta)$ may exhibit different decay behavior, but will still vanish as $N \to \infty$. See also  \citet{RajanEtAl2013}. 

\section{Appendix B: DPI-satisfying measures} 

DPI is satisfied by all measures of the $F$-information form \citep{CsiszarShields2004, KinneyAtwal2013},
\begin{eqnarray}
I_{F}[M;R] \equiv \int dR\, dM\, p(R) p(M) F \left( \frac{p(R,M)}{p(R)p(M)} \right),
\end{eqnarray}
where $F(x)$ is a convex function for $x \ge 0$. Mutual information corresponds to $F(x) = x \log x$ whereas $F(x) = (x^{\alpha} - 1)/(\alpha-1) $ yields a more general ``R\'{e}nyi information'' measure \citep{Renyi1961} that reduces to mutual information when $\alpha = 1$.  DPI-satisfying measures other than mutual information have been used for filter inference in a number of works, including \citet{Paninski2003} and \citet{KouhSharpee2009}. A discussion of the differences between DPI-satisfying measures and some non-DPI-satisfying measures can be found in \citep{KinneyAtwal2013}.

\section{Appendix C: DPI-optimality}

Assume $\theta < \theta^*$ by Eq.\ \ref{eq:strong}. Because $R \leftrightarrow R^* \leftrightarrow M$ is a Markov chain, the KL divergence between $p(R^*|R,M)$ and $p(R^*|R)$ can be decomposed as $D( p(R^*|R,M) || p(R^*|R)) = I[R^*;M] - I[R;M].$ If this quantity is zero, then $R^* \leftrightarrow R \leftrightarrow M$ is also Markov chain, implying $\theta^* \leq \theta$, a contradiction. This KL divergence must therefore be positive, i.e.\ $I(\theta) < I(\theta^*)$. So if $\theta^* \in \Theta_{\rm DPI}$, then for every $\theta \not \in \Theta_{\rm DPI}$, $\theta \not \in \Theta_I$ as well. This proves $\Theta_I = \Theta_{\rm DPI}$. 

\section{Appendix D: information equivalence}

First we observe that if $\theta_1$ and $\theta_2$ make isomorphic predictions then they are information equivalent. This is readily shown from the fact that $\mathcal{D}[R;M]$ is invariant under arbitrary invertible transformations of $R$ \citep{KinneyAtwal2013}. Next we show the converse: if $\theta_1$ and $\theta_2$ are information equivalent, the predictions $R_1$ and $R_2$ must be isomorphic. Here is the proof. If $\theta_1 \simeq \theta_2$, then $\mathcal{D}[R_1;M]= \mathcal{D}[R_2;M]$ for all $\mathcal{D}$, and in particular $I[R_1;M] = I[R_2;M]$. In Appendix C we showed that $I[R;M] = I[R^*;M]$ implies $R^* \leftrightarrow R \leftrightarrow M$ is a Markov chain. Imagining an SRM experiment in which $\theta^* = \theta_1$ and $\pi(M|R) = \delta(M - R)$, we find that $R_1 \leftrightarrow R_2 \leftrightarrow R_1$ is a Markov chain. This implies the mapping $R_2 \to R_1$ is one-to-one. Similarly, $R_1 \to R_2$ is one-to-one. $R_1$ and $R_2$ are therefore bijective.  

\subsection*{Acknowledgments}
We thank William Bialek, Curtis Callan, Bud Mishra, Swagatam Mukhopadhyay, Anand Murugan, Michael Schatz, Bruce Stillman, and Ga\v{s}per Tka\v{c}ik for helpful conversations. Support for this project was provided by the Simons Center for Quantitative Biology at Cold Spring Harbor Laboratory.

\clearpage

\begin{figure}[h]
\hfill
\begin{center}
\includegraphics{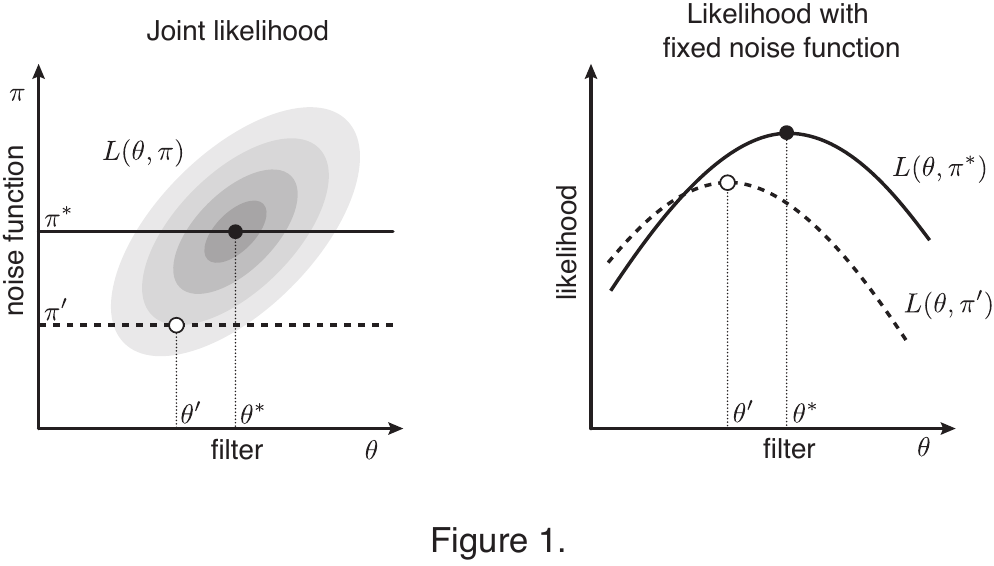}
\end{center}
\label{fig:likelihood}
\end{figure}

{\bf Figure 1}: Maximizing likelihood with an incorrect noise function will generally bias the inferred filter. The per-datum log likelihood $L(\theta,\pi)$ will typically depend on both the filter $\theta$ and the noise function $\pi$ in a correlated manner (left panel). Values of a schematic $L(\theta,\pi)$ are illustrated in gray, with darker shades indicating larger likelihood. If the correct noise function $\pi^*$ is assumed (solid line), maximizing $L(\theta,\pi^*)$ will yield the correct filter $\theta^*$ (filled dot). However, if an incorrect noise function $\pi'$ is assumed (dashed line), maximizing $L(\theta,\pi')$ will typically lead to an incorrect filter $\theta'$ (open dot). \\

\clearpage

\begin{figure}[h]
\hfill
\begin{center}
\includegraphics{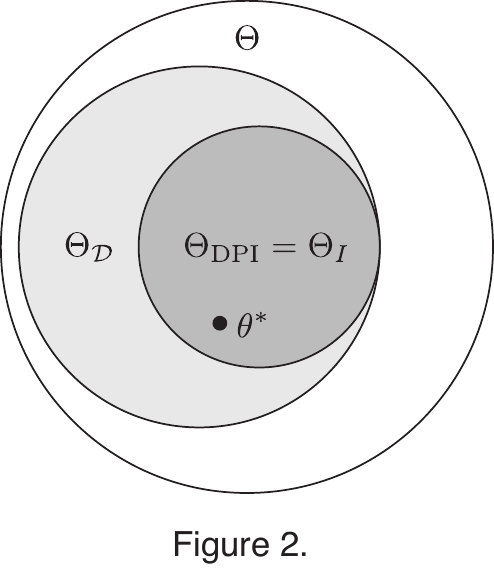}
\end{center}
\label{fig:venn}
\end{figure}

{\bf Figure 2}: Venn diagram illustrating filter sets maximizing different DPI-satisfying measures. In general, different DPI-satisfying dependence measures, e.g.\ mutual information $I$ and some other measure $\mathcal{D}$, will be maximized by different sets of filters, respectively represented here by $\Theta_I$ and $\Theta_{\mathcal{D}}$.  $\Theta_{\rm DPI}$ is the intersection of the optimal sets of all such DPI-satisfying measures. Mutual information has the important property that $\Theta_I = \Theta_{\rm DPI}$ whenever $\theta^* \in \Theta$; this is not true of all DPI-satisfying measures. \\

\clearpage

\begin{figure}[h]
\hfill 
\begin{center}
\includegraphics{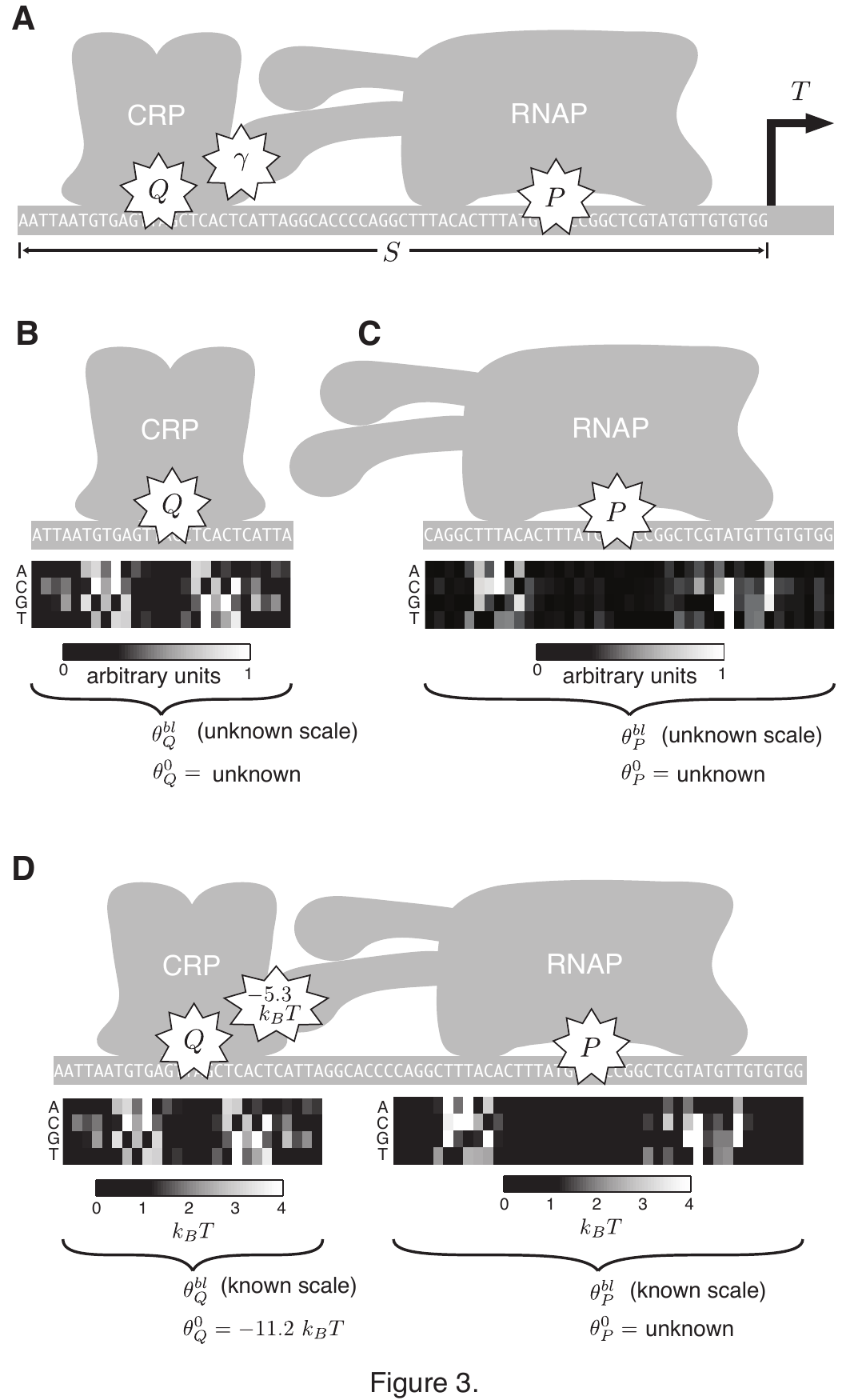}
\end{center}
\label{fig:promoter}
\end{figure}

\clearpage

{\bf Figure 3}: A linear-nonlinear filter modeling the biophysics of transcriptional regulation at the \emph{Escherichia coli} \emph{lac} promoter. (A) The biophysical model inferred by \citet{KinneyEtAl2010} from Sort-Seq data. Each signal $S$ is a 75 bp DNA sequence differing from the wildtype \emph{lac} promoter by $\sim 9$ randomly scattered substitution mutations. $Q$ and $P$ denote the sequence-dependent binding energies of the proteins CRP and RNAP to their respective sites on this sequence $S$; both $Q$ and $P$ were modeled as linear filters of $S$. $\gamma$ is a sequence-independent interaction energy between CRP and RNAP. The resulting transcription rate $T$, of which the Sort-Seq assay produces noisy measurements $M$, is assumed to depend on $Q$, $P$, and $\gamma$ in a specific nonlinear manner dictated by the hypothesized biophysical mechanism (Eq.\ \ref{eq:kinney_model}; all energies are in units of $k_B T$). (B) The linear filter $Q$ is defined by parameters $\theta^{bl}_Q$ and $\theta^0_Q$ via Eq.\ \ref{eq:linear_filters}. Inferring these parameters by maximizing the mutual information $I[Q;M]$ determines $\theta^{bl}_Q$ up to an unknown scale and leaves $\theta^0_Q$ undetermined. (C) Analogous results are obtained for the parameters $\theta^{bl}_P$ and $\theta^0_P$ when $I[P;M]$ is maximized. (D) Because of the inherent nonlinearity in Eq.\ \ref{eq:kinney_model} (right-hand side), maximizing $I[T;M]$ breaks diffeomorphic modes, fixing the values of $\theta^{bl}_Q$, $\theta^{bl}_P$, and $\theta^0_Q$ in units of $k_B T$. The parameter $\theta^0_P$ remains undetermined. \\

\clearpage

\begin{figure}[h]
\hfill 
\begin{center}
\includegraphics{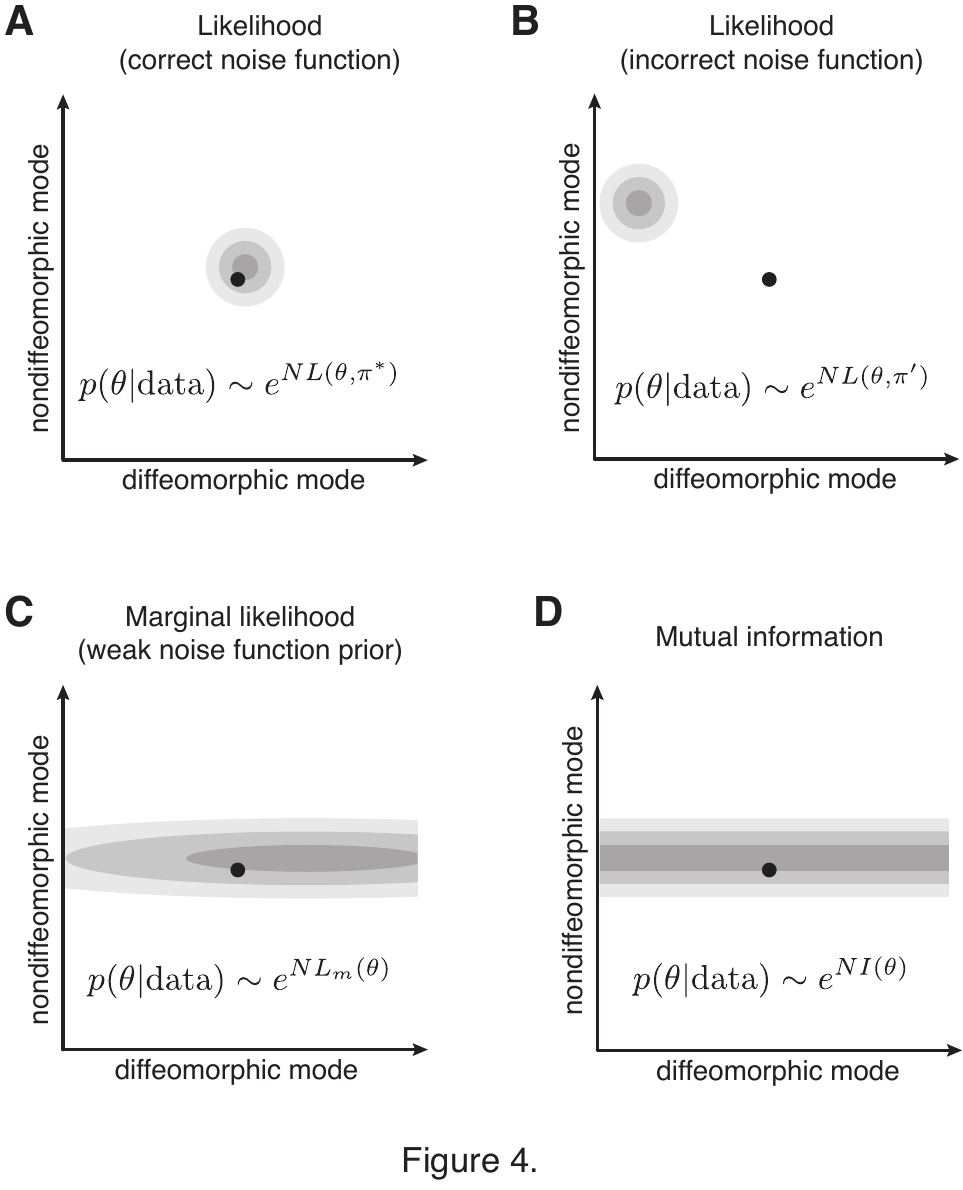}
\end{center}
\label{fig:hessians}
\end{figure}

{\bf Figure 4}: Schematic illustration of constraints placed on diffeomorphic and nondiffeomorphic modes by different objective functions. The dot in each panel represents the correct filter $\theta^*$; shades of gray represent the posterior distribution $p(\theta | {\rm data})$. (A,B) Likelihood (Eq.\ \ref{eq:likelihood}) places tight constraints (scaling as $N^{-1/2}$ as $N \to \infty$) along both diffeomorphic and nondiffeomorphic modes. (A) $\theta^*$ will typically lie within error bars if the correct noise function $\pi^*$ is used. (B) However, if an incorrect noise function $\pi'$ is used, $\theta^*$ will generally violate inferred constraints along both diffeomorphic and nondiffeomorphic modes. (C) Marginal likelihood (Eq.\ \ref{eq:marginal_likelihood}) computed using a sufficiently weak prior $p(\pi)$ will place tight constraints on nondiffeomorphic modes and weak constraints (scaling as $N^0$ as $N \to \infty$) along diffeomorphic modes. (D) Mutual information (Eq.\ \ref{eq:mi}) places tight constraints on nondiffeomorphic modes but provides no constraints whatsoever on diffeomorphic modes. 

\end{document}